# DATA JUSTICE IN DIGITAL SOCIAL WELFARE: A STUDY OF THE RYTHU BHAROSA SCHEME

Silvia Masiero, University of Oslo, silvima@ifi.uio.no

Chakradhar Buddha, LibTech India, samalochana@gmail.com

**Abstract:** While digital social protection systems have been claimed to bring efficacy in user identification and entitlement assignation, their data justice implications have been questioned. In particular, the delivery of subsidies based on biometric identification has been found to magnify exclusions, imply informational asymmetries, and reproduce policy structures that negatively affect recipients. In this paper, we use a data justice lens to study Rythu Bharosa, a social welfare scheme targeting farmers in the Andhra Pradesh state of India. While coverage of the scheme in terms of number of recipients is reportedly high, our fieldwork revealed three forms of data justice to be monitored for intended recipients. A first form is *design-related*, as mismatches of recipients with their registered biometric credentials and bank account details are associated to denial of subsidies. A second form is *informational*, as users who do not receive subsidies are often not informed of the reason why it is so, or of the grievance redressal processes available to them. To these dimensions our data add a *structural* one, centred on the conditionality of subsidy to approval by landowners, which forces tenant farmers to request a type of landowner consent that reproduces existing patterns of class and caste subordination. Identifying such data justice issues, the paper adds to problematisations of digital social welfare systems, contributing a structural dimension to studies of data justice in digital social protection.

## 1. INTRODUCTION

Over the last decade, a substantial literature has related digital identification of users to efficacy and justice in social protection systems (cf. Gelb & Clark, 2013; World Bank, 2016; Gelb & Metz, 2018). Taylor (2017) defines data justice as "fairness in the way people are made visible, represented and treated as a result of their production of digital data". Visualisation, representation and treatment of people through digital data production acquire greater relevance in contexts of vulnerability, such as informal labour (Krishna, 2021), labour in the gig economy (Heeks et al., 2020), forced migration and displacement (Cheesman, 2020; Martin & Taylor, 2021). Overlap of data injustices with pre-existing forms of vulnerability has been highlighted as a pressing concern on the data justice agenda (Heeks & Renken, 2018; Dencik et al., 2019; Heeks & Shekhar, 2019).

Yet, concern has emerged on the extent to which digital and, especially, biometric identification can concur to data justice in social protection systems. From the legal perspective, subordination of essential rights to enrolment in biometric databases (Srinivasan et al., 2018) makes enrolment a condition for rights that are, on paper, universal. From the informational point of view, uncertainty remains on the treatment of user data, leading research to question the extent to which coercive enrolment results in transparent data handling (Cerna Aragon, 2021). As a result, research finds a space of enquiry in the extent to which digital identification can provide welfare to impoverished people, and on the justice concerns with such a process.

We engage the problem with a study of Rythu Bharosa, a social welfare scheme announced in 2019 to provide cash subsidies to farmers in the Indian state of Andhra Pradesh. We use a data justice lens to ask, *how does digital social protection meet the needs of vulnerable beneficiaries?* Drawing on narratives collected from users entitled to Rythu Bharosa in early 2020, our analysis identifies two dimensions of data injustice (design-related and informational) in line with Masiero and Das (2019), but builds on data in finding a structural dimension, mirroring the concept of structural data justice in Heeks and Shekhar (2019), that reflects the perpetuation of subordination of tenant farmers to landowners in the digital scheme.

As a result, the paper contributes a structural dimension to studies of data justice in digital social protection. Heeks and Shekhar (2019: 994) define structural data justice as "the degree to which the interests and power in wider society support fair outcomes in other forms of data justice". Conceptualising structural data justice in digital social protection allows visualising perpetuation of exploitative class and caste relations in datafied schemes and the way technology participates in these, illuminating its implications for vulnerable users.

The paper is structured as follows. We first summarise the literature on data justice in social protection schemes, highlighting the orthodoxy of digital identity for development and the main critiques to it. We then introduce Rythu Bharosa and its digital architecture, illuminating the data justice lens through which we study it. In the analysis, we outline three forms of data justice in digital social protection: two, *design-related* and *informational*, are derived from Masiero and Das (2019), while a third, *structural*, stems from our data on Rythu Bharosa. The discussion positions our contribution in data justice research, highlighting its implications for digital social protection.

## 2. LITERATURE AND CONCEPTUAL BACKGROUND

A focus on visualisation, representation and treatment of people through data defines data justice (Taylor, 2017). The intersection of data injustices with pre-existing forms of social and economic vulnerability is highlighted on data justice research agendas (Heeks & Renken, 2018; Dencik et al., 2019; Heeks & Shekhar, 2019), inspiring a focus of data justice on social protection schemes.

Social protection is defined as "public and private initiatives that provide income or consumption transfers to the poor, protect the vulnerable against livelihood risks, and enhance the social status and rights of the marginalised; with the overall objective of reducing the economic and social vulnerability of poor, vulnerable and marginalised groups" (Devereux & Sabates-Wheeler, 2004). Among extant domains of data justice research, social protection offers a space where data-based architectures promise better functioning of existing programmes (Masiero & Shakthi, 2020). Such a promise suggests that, by matching people's biometric credentials with their entitlements, more accurate delivery of programmes can be achieved. Digital identification and delivery of entitlements combat both inclusion and exclusion errors, ensuring that leakage is tackled and that those entitled are able to receive their benefits (cf. Gelb & Metz, 2018).

Masiero and Bailur (2021) propose a further taxonomy of the theoretical link of digital identity and development, finding three channels along which such a taxonomy is articulated. First, a link between digital identity and the provision of improved public services is theorised. The core idea is that better identification of users, through the improvement of targeting, leads to exclusion of non-entitled users from benefits, with a positive effect on reduction of leakage and diversion from social protection schemes (Gelb & Clark, 2013; World Bank, 2016; Gelb & Metz, 2018). Made explicit in the global agenda on digital identity for development, the assumption of digital identity leading to better service provision finds, however, only limited empirical confirmation (Bhatia et al., 2021; Carswell & De Neve, 2021; Chaudhuri, 2021; Martin & Taylor, 2021).

A second channel through which digital identity is supposed to foster development pertains to ability to improve assistance to systemically excluded minorities. This point applies to refugees and internally displaced people, for whom it is suggested that a digital identification system can guarantee registration even in conditions of statelessness. The promise, as summarised by the World Bank (2016), is that of recognition of identity where one is not easily established. From a theoretical perspective, the inclusion of minorities transcends the promise of better services, as it seeks to guarantee solutions that combat the roots of exclusion (Bhatia et al., 2021).

Thirdly, improvement of humanitarian assistance is a further promise held by digital identity. The humanitarian sector is becoming based on recognition of beneficiaries that is digitally grounded, conceived and operated (Donovan, 2015). The digitisation of the humanitarian sector informs a new focus of digital identity in aid (Weitzberg et al., 2021), as abilities to recognise entitled subjects produce systems that can assign assistance to the needful.

Articulated along the lines of improved public services, inclusion of minorities and better delivery of humanitarianism, the link of digital identity and development is widely recognised by advocates of digital provision of social protection. The same link is, however, problematised in all its three components from data justice and social protection research (Masiero & Bailur, 2021). Below we engage the main problematisations of each component of the link in point.

First, the link between digital identity and improved public services is questioned especially by studies of user experience of digital social protection. Exclusions from biometrically-mediated services are proved, in the case of large food security schemes such as India's Public Distribution System, both in quantitative (Muralidharan et al., 2020) and qualitative terms (Hundal et al., 2020; Chaudhuri, 2021). Experiences of digital access to the state lead to subordination of fundamental rights to biometric recognition, leading to trading of personal data for services (Srinivasan et al., 2018). studies of digital social protection during COVID-19 illustrate the exclusions of social protection users in times of global emergency (Cerna Aragon, 2021; López, 2021): exclusions are harder to measure for groups, such as migrants and displaced populations, which may not be seen and counted in mainstream statistics (de Souza, 2020; Milan & Treré, 2020).

Second, the link of digital identity to inclusion of minorities is problematised as well. In a study of informal workers in India, Krishna (2021) finds a systematic discrepancy between promotion of digital identity as an enhancer of protection and the reality of workers, subject to surveillance rather than assistance. Studies of refugees being registered through digital identity schemes find similar forms of surveillance: in the refugee stories from Jordan, Lebanon and Uganda collected by Schoemaker et al. (2021), digital identity results in a blurred boundary between assistance and policing. In the cases of Uganda and Bangladesh studied by Martin and Taylor (2021), systematic differences are found in regulatory practices: while Uganda has enacted a virtuous cycle of assistance, narratives from Bangladesh still reveal how digital technology acts as an enhancer of discriminatory practices towards refugee populations.

Completing these narratives, studies of digital identity in aid reveal the coexistence of logics of recognition and surveillance (Weitzberg et al., 2021). Cheesman (2020) studies the ambivalence of decentralised technologies of self-sovereign identity, noting the limits to the thesis that views these as "emancipatory" architectures for refugees. In his study of biometric registration in one of the largest refugee camps of Kenya, Iazzolino (2021) finds a duality of assistance and repression in biometric systems for refugees, for whom profiling has direct effects on the ability of the state to monitor and police their movements. As observed in Donovan (2015), this duality is implicit in digital monitoring, resulting in exacerbated effects for populations at risk during displacement (Iazzolino, 2021). Problematising the ability of digital technologies to offer better assistance to refugees, digital identity arises again as questionable in its ability to strengthen social protection.

It is against this backdrop that we set to study how digital social protection meets the needs of vulnerable beneficiaries. From the voices of users of Rythu Bharosa, a subsidy scheme designed for vulnerable farmers in the southern Indian state of Andhra Pradesh, we set to explore whether, and how, digital social protection schemes can fill existing exclusions and ensure data justice to beneficiaries. From Rythu Bharosa recipient narratives we derive more problematisations of the link between digital identity and development.

## 3. METHODOLOGY

Case selection for this paper was based on considerations of timeliness and conformity of the Rythu Bharosa scheme with the research object, digital social protection, on which our question is centred. Time wise, the implementation of Rythu Bharosa from October 2019 made it possible to gather first-hand information on the scheme and have one of the authors establish access to recipients, with whom group discussions were conducted in the aftermath of the rollout. Access was established under the aegis of an Indian collective of engineers, social workers and social scientists, inspired by the Right to Information Movement (Sharma, 2013) and working to promote democratic engagement in rural public service delivery.

Group discussions with users entitled to Rythu Bharosa revolved on three core themes. A first theme was centred on *access* to the scheme: questions on this topic regarded the ability, or lack thereof, of users to receive their entitlements. In the cases where users had their entitlements denied, questions invited narratives on the reasons offered by officials at bank and village levels on such an outcome, and on redressal mechanisms sought by users. A second theme was centred on the *conditions* surrounding entitlement, and invited users to expand on the conditions to apply for Rythu Bharosa and their enforcement through the digital system. Finally, a third set of questions pertained to *information* received on the scheme, asking how such information was conveyed through the digital system and its ability to enable the performance of applications.

| Date       | Interviewees                | No. |
|------------|-----------------------------|-----|
| 28.12.2019 | Farmers                     | 5   |
| 29.12.2019 | Agriculture union activist  | 1   |
| 29.12.2019 | Tenant farmers              | 6   |
| 30.12.2019 | Farmers                     | 6   |
| 31.12.2019 | Government officials        | 2   |
| 03.01.2020 | Tenant farmers              | 8   |
|            | Tenant farmers              | 6   |

| 04.01.2020 | Tribal farmers | 4 |
|---|---|---|
| 05.01.2020 | Village volunteer | 1 |
|  | Women tenant farmers | 3 |
|  | Village volunteer | 1 |
|  | Tenant farmers | 4 |
| 06.01.2020 | Political party member | 1 |
| Total |  | 47 |

Table 1: Summary of Interviews

Field data were progressively integrated with statistics on the Rythu Bharosa scheme, as well as press releases and state government documentation on the programme. While limited to the immediate aftermath of implementation, group discussions with entitled beneficiaries have been key to illuminating user appraisal of a large digital social protection programme, as well as its affordances and challenges for farmers across gender and caste.

## 4. CASE BACKGROUND

After a landslide victory in the State Assembly elections of 2019, the newly elected government of the Indian state of Andhra Pradesh announced a set of social welfare schemes termed Navaratnalu (meaning "nine gems"). Navaratnalu is a set of nine social welfare schemes introduced with the objective of bringing positive change in the lives of underprivileged and vulnerable groups. Navaratnalu follows an integrated approach including schemes on agriculture, health, education and housing among other sectors.

The Rythu Bharosa scheme, one of the nine schemes within Navaratnalu, aims to support small and marginal farmers in the state of Andhra Pradesh. The state government implemented the scheme from October 15th, 2019, for augmenting the income of landowner farmer families and to the Scheduled Caste, Scheduled Tribe, Backward Caste and Minority Landless Tenant farmer families by providing financial assistance across the state.[1] The scheme consists in the provision of a cash subsidy to landowner farmers, in order to meet investments (especially high during the crop season) and thereby mitigate their financial struggles.

As part of the scheme, farmers identified as entitled receive 13,500/- Indian rupees per year. Out of these, the Central Government transfers 6,000/- across 3 equal installments in the name of PM Kisan, a central government scheme aimed to extend income support to all landholding farmers' families in the country having cultivable land. PM Kisan is a Government of India scheme that guarantees

---

[1] https://pmmodiyojana.in/ysr-rythu-bharosa-list/, accessed 14 May 2021.

Rs 6,000 per year to small and marginal farmers as minimum income support.[2] The state government completes this transferring 7,500/- in two installments. Eligibility criteria vary slightly across users: all the farmers who received land entitlements under the Record of Forest Rights (ROFR), a scheme that defends the rights of forest cultivating communities, and tenant farmers are ineligible for the PM Kisan scheme, in these cases the state government pays the whole 13,500/- entirely from its budget.

Rythu Bharosa ensures delivery of subsidy to entitled users by a management information system (MIS) that matches user identity with records of their entitlement. User identity is recorded with Aadhaar, the database that records biometric credentials of Indian residents and assigns them a 12-digit number. The Rythu Bharosa information system matches Aadhaar credentials with the status assigned to users, also linking to each record the bank account where the subsidy money is to be deposited. Based on a system of digital identification of users, the scheme offers a paradigmatic case of the object, digital social protection, that our research aims to investigate.

## 5. FINDINGS

At the time of writing, about 51 lakh (5.1 million) farmers are getting their benefits from Rythu Bharosa. This coverage rate testifies to the wide reach of the programme, and to its ability to deal with issues suffered by farmers subjected to income limitations.

Farmer narratives, however, revealed multiple dimensions questioning principles of data justice in the programme. Such dimensions have informational, design-related and structural aspects.

### 5.1. Informational data justice

Masiero and Das (2019) refer to informational data justice as a situation in which users of social protection are properly informed on how their data are treated, and of the decisions made on the basis of such data. Narratives collected from recipients entitled to receive Rythu Bharosa have, however, revealed several issues with this dimension of data justice.

- *Awareness of entitlements*. Group discussions with beneficiaries reveal that the assignment of entitlements under Rythu Bharosa is not always clear, and is in many cases opaque. To determine entitlement, an existing list - available to the State Government from extant farmer databases - has been augmented with the request to village volunteers to identify eligible farmers in given areas. Village volunteers are the street level bureaucrats whose job is to help people in getting their entitlements. They collect requisite documents from people and deposit them in village secretariats which are administrative offices at panchayat level. The digital trial of the beneficiaries start from there. Yet, at the moment of claiming Rythu Bharosa subsidies, several farmers found themselves excluded from such a list. Some farmers suggest this may be due to errors in the praja sadhikara survey, which took place in 2016 and was supposed to capture the socio-economic status of all households in the state.

- *Mismatches with existing Aadhaar seedings*. Way back in 2014 the state government seeded all the land accounts in the state with Aadhar and the local officials were given unreasonably short time to complete the process. To avoid being pointed out for not meeting deadlines they seeded land accounts with whichever Aadhar numbers they came across. Shortly before the rollout of Rythu Bharosa, the local government was required to do verification and correction of field mapping of all Aadhaar numbers of beneficiaries in 15 days. This was found, some farmers suggest, a challenging task given the density of households to cover. In fact, several farmers report erroneous Aadhaar seedings, resulting in entitlements being deposited in bank accounts that do not belong to them. A redressal issue is also reported, since discussions with farmers highlight their inability to verify online the Aadhaar number that their land account number is seeded with. As stated by a farmer in one of the village areas, information is "hidden behind the logins": in the way the MIS is planned out, even the officers responsible for it find it difficult to investigate missing subsidies.

---

[2] https://www.business-standard.com/about/what-is-pm-kisan-yojana, accessed 14 May 2021.

- *Unclear reasons for rejection*. When farmers apply for Rythu Bharosa, rejection can happen at two stages. The system distinguishes between *bank rejection* and *village rejection*: in the former case, subsidies are rejected at the level of the bank, due to data mismatches or absence of data for the individual in point. In the latter case, subsidies are rejected at the village level, where local databases have either not been accurately seeded with user data, or not seeded at all or rejected due to ineligibility. Such information is however not available to the user, who is therefore left with uncertainty on the reason why their subsidy has been rejected.

- *Lack of effective grievance redressal mechanisms for farmers*. Our fieldwork did not show evidence of any form of effective grievance redressal for the farmers that, while entitled to Rythu Bharosa, still received rejections either at the bank or at the village level. The system is predicated on the idea that grievance redressal is needed in order to assist beneficiaries for whom entitlement is not converted into reality. At the same time, our fieldwork gave no evidence of any attempt of the state to engage in grievance redressal, which is especially problematic in the context of repeated rejections of effectively entitled beneficiaries.

All these points suggest issues of informational justice, since treatment of user data and decisions made on their basis are often unclear. Mismatches between Aadhaar records, landholder and bank accounts are difficult for both users and officers to investigate, making it difficult for users to obtain the information needed on their cases.

**5.2. Design-related data justice**

Masiero and Das (2019) conceive design-related data justice as a situation in which the design of social protection schemes is in line with the needs that users experience. This concept follows the idea of design-reality gaps as in Heeks (2002), where ICT4D project failure may result from gaps between the way in which the technology is designed and the reality perceived by recipients. In our fieldwork, several questions of design-related data justice have emerged.

- *System design preventing change of wrong records*. In case of wrong digital records, users who have been denied subsidies can request a change. Many farmers and lower level officials complained that changes made in the digital land records website, hasn't got reflected in the Rythu Bharosa website thereby denying entitlements to a large number of farmers. But there have been cases where the clerk has printed out the old record, and justified that by stating "this was set in this way" – preventing changes from being done. Viewed from user voices, the MIS seems exceptionally difficult when it comes to make changes to existing records, changes that may be determinant for someone's ability to effectively receive the subsidy.

- *Complexity of system update procedures*. Related to the above, farmers who want to update their records have to approach the village volunteer or village revenue officer, asking for updates to be done. Again, if updates are not done in time, entitled users can be excluded from the system.

In the case of tribal users, knowing whether the money has arrived or not means multiple trips to the bank: it can occur that they travel all the way to the bank to find that the money is not there.

- *Uncertainty of records*. The system is built in such a way that the user should know how much has been paid, but a common thread across our group discussions is that transparency is not a characteristic of the system. Respondents from a tribal village that we visited in January 2020 define this as "the most opaque scheme" they have dealt with, as the whole process of assignation of entitlements is perceived to be hidden to the user. Even when going to the village volunteer, the system is designed in such a way that the volunteer is not enabled to have the information needed to help. Tribal areas have the additional problem that forest lands are not registered in the land record system, resulting in information on such lands not being available in the public domain. The system, users told us, is like a lottery: even those who do get subsidies often do not know why they got it.

To sum up, several design-reality gaps exist between the design of Rythu Bharosa and the lives of recipients. These revolve around system design preventing the change of wrong records, the complexity of system update procedures and uncertainty of the records entered in the system.

## 5.3 Structural data justice

Our analysis finds an additional form of injustice, which pertains to the MIS reflecting structural, long-term relations of class and caste in landowning. Instances of such a type of data injustice are presented here, observing how technology intersects with existing power relations.

- *Perpetuation of class relations.* Tenant farmers need a signature from landowners to obtain subsidy from the Rythu Bharosa scheme, a signature without which the application for the subsidy is incomplete. Village volunteers ensure that families get state benefits, but in the absence of the landowners' signature, it is not possible to give such an assurance.

  In practice, tenant farmers find themselves in the situation that the landowner can just decide not to sign the agreement entitling them to Rythu Bharosa, which perpetuates extant forms of tenant farmer subordination. As explained by one of the village volunteers, the idea that the state government promoted through Rythu Bharosa is to invite landowners to pursue some form of legal agreement with tenant farmers, but limited evidence of agreements taking place has emerged from our discussions with users. In one of the group discussions, we became aware of landowner farmers applying for the subsidy on a certain land, without informing their tenants who hence became not entitled to the same.

- *Perpetuation of caste relations*. The role of caste in landowning relations has been evidenced in our field data. In an interview with Dalit women in one of the villages, a specific caste has emerged as the landowning caste in the area, with farmers reporting having tilled the same land from the same landowner for over twenty years. But even then, they were given no agreement to receive Rythu Bharosa entitlement. Some of them did not do so as they were not informed of the requirement for landowner signature, which is however embedded in the MIS.

  In discussions with the same tenant farmers, it has emerged that in some cases, tenants are reluctant to ask a signature from their landowners even if they are aware of the requirement. Two women farmers in the group said that, as they had worked in the land for their landowners for over twenty years, they felt that requesting a signature for access to Rythu Bharosa could upset the trust relation built with landowners. As powerfully summarised by one of them: "we have been trusted all these years, how would we put that in peril?"

As a result, the Rythu Bharosa MIS feeds into a history of power relations between farmers and landowners. Rythu Bharosa is a subsidy of the state to help needful farmers, but if that subsidy is (a) conditional to approval by landowners, (b) sent to the wrong account, or (c) accompanied by no information on rejection reasons, data justice issues of informational, design-related and structural nature are likely to persist.

# 6. DISCUSSION

Faced with the orthodoxy of digital identity for development and its critiques, in this paper we set to investigate how digital social protection meets the needs of vulnerable beneficiaries. Our study revealed that Rythu Bharosa is a well-planned scheme to help vulnerable farmers, which however reveals several aspects of data justice to monitor. These belong to the design-related and informational domain, theorised in Masiero and Das (2019), and to the domain of structural data justice as theorised in Heeks and Shekhar (2019). We suggest that data injustices of the structural type reinforce adverse power relations, such as the dependency of farmers on landowners, that hamper the economic empowerment that social protection is supposed to enable.

The paper contributes a taxonomy of forms of data justice in digital social protection, which enriches existing ones (Masiero & Das, 2019) and adds a structural dimension to them. As noted in Masiero and Bailur (2021), the orthodoxy of digital identity as a route to development is questioned on theoretical and empirical grounds, which makes it important to assess the extent to which it comes alive in existing social protection programmes. Schemes like Rythu Bharosa are paradigmatic of the state's commitment to vulnerable residents, and their provisions embody the strengths of social protection as defined by Devereux and Sabates-Wheeler (2004). Focusing on the MIS that operates the scheme, our study affords understanding the dimensions of data justice that need monitoring.

The study reveals dimensions of design-related and informational justice, that Masiero and Das (2019) highlight in digital social protection. From a design-related perspective, the system makes it difficult to perform operations, such as amending user records and connecting them to bank accounts, that are needed for subsidies to be credited. From an informational perspective, the programme emerges as especially opaque: users, it emerges, are extremely uncertain on the reasons behind acceptance or rejection of their subsidy application. Chaudhuri (2021) points at this specific dimension: an "opaque" state, that does not reveal the reasons for key decisions on social protection, is extremely problematic for its residents.

To these dimensions, the study adds that of structural data justice. Such a dimension impinges on power relations as in Heeks and Shekhar (2019) but reflects such relations in the lived reality of users of digital social protection. Echoing studies of colonial relations embedded in biometric systems (Weitzberg, 2020), our study shows that the Rythu Bharosa MIS embodies extant relations of subordination between tenant and landowner farmers, whose signature is requested when tenants wish to apply for the subsidy. Doing so, the system transposes into data the power relations of the real world, hampering the economic empowerment that the system could ensure.

Structural data justice is especially important in the current context of digital social protection. With our fieldwork conducted ahead of the declaration of COVID-19 as a pandemic, our study meets a world in which digital social protection is adapting to new needs (Masiero, 2020; Milan et al., 2021). Across countries, the COVID-19 pandemic has made effective social protection systems a priority, posing new data justice concerns that overlap with the previous ones (Taylor et al., 2020; Martin, 2021). In a world where social protection has received a new, suddenly enhanced importance, it has become crucial to find ways for social protection to tackle structural power imbalances, acting on cases in which such imbalances are digitally reinforced.

Finally, a limitation of the study relates to the temporal window of late 2019-early 2020 in which primary data are concentrated. While the ongoing pandemic has largely changed the way fieldwork is conducted, we find value in our data due to their ability to voice user perspectives, putting them in the context of a digital social protection scheme with a wide coverage. The pandemic has maximised the importance of strong social protection, making our results especially important in a context of continued global emergency (Milan et al., 2021). As the consequences of COVID-19 on vulnerable groups perdurate, we hope that this study enables the recognition of core dimensions of data justice in digital social protection.